\documentclass[12pt,preprint]{aastex}

\slugcomment{Accepted to the Astrophysical Journal Supplement Series}

\newcommand{\GPB}{\mbox{GP-B}}
\newcommand{\IM}{\mbox{IM~Peg}}
\newcommand{\Msol}{\mbox{M\raisebox{-.6ex}{$\odot$}}}
\newcommand{\Rsol}{\mbox{R\raisebox{-.6ex}{$\odot$}}}

\defcitealias{GPB-I}{Paper I}
\defcitealias{GPB-II}{Paper II}
\defcitealias{GPB-III}{Paper III}
\defcitealias{GPB-IV}{Paper IV}
\defcitealias{GPB-V}{Paper V}
\defcitealias{GPB-VI}{Paper VI}
\defcitealias{GPB-VII}{Paper VII}

\shorttitle{VLBI for {\em GP-B}. VI.}
\shortauthors{Ransom et al.}

\begin{document}

\title{VLBI for {\em Gravity Probe B}. VI. The Orbit of IM Pegasi and
the Location of the Source of Radio Emission}

\author{R. R. Ransom\altaffilmark{1,2}, N. Bartel\altaffilmark{1},
M. F. Bietenholz\altaffilmark{1,3}, D. E. Lebach\altaffilmark{4},
J.-F. Lestrade\altaffilmark{5}, M. I. Ratner\altaffilmark{4}, and
I. I. Shapiro\altaffilmark{4}}

\altaffiltext{1}{Department of Physics and Astronomy, York University,
4700 Keele Street, Toronto, ON M3J 1P3, Canada}

\altaffiltext{2}{Now at Okanagan College, 583 Duncan Avenue West,
Penticton, B.C., V2A 2K8 and also at the National Research Council of
Canada, Herzberg Institute of Astrophysics, Dominion Radio
Astrophysical Observatory, P.O. Box 248, Penticton, B.C., V2A 6K3,
Canada}

\altaffiltext{3}{Now also at Hartebeesthoek Radio Astronomy Observatory,
PO Box 443, Krugersdorp 1740, South Africa}

\altaffiltext{4}{Harvard-Smithsonian Center for Astrophysics, 60
Garden Street, Cambridge, MA 02138, USA}

\altaffiltext{5}{Observatoire de Paris/LERMA, 77 av. Denfert Rochereau, F-75014 Paris, France}


\begin{abstract}

We present a physical interpretation for the locations of the sources
of radio emission in IM~Pegasi (IM~Peg, HR~8703), the guide star for
the NASA/Stanford relativity mission Gravity Probe B.  This emission
is seen in each of our 35 epochs of 8.4-GHz VLBI observations taken
from 1997 to 2005\@.  We found that the mean position of the radio
emission is at or near the projected center of the primary to within
about $27$\% of its radius, identifying this active star as the radio
emitter. The positions of the radio brightness peaks are scattered
across the disk of the primary and slightly beyond, preferentially
along an axis with position angle, p.a.\ = $-38\arcdeg\pm8\arcdeg$,
which is closely aligned with the sky projections of the orbit normal
(p.a.\ = $-49.5\arcdeg\pm8.6\arcdeg$) and the expected spin axis of
the primary. Comparison with simulations suggests that brightness
peaks are $3.6^{+0.4}_{-0.7}$ times more likely to occur (per unit
surface area) near the pole regions of the primary (latitude,
$|\lambda| \geq 70\arcdeg$) than near the equator ($|\lambda| \leq
20\arcdeg$), and to also occur close to the surface with $\sim$2/3 of
them at altitudes not higher than 25\% of the radius of the primary.

\end{abstract}

\section{Introduction}

IM~Pegasi (\IM; HR~8703; HD~216489; FK5~3829) is the radio-bright
binary star which served as the guide star for the Gravity Probe B
(\GPB) mission, the spaceborne relativity experiment developed by NASA
and Stanford University to test two predictions of general relativity
(GR).  This paper is the sixth in a series of seven describing the
program of very-long-baseline interferometry (VLBI) undertaken in
support of \GPB\@.  In the first paper in the series we give an
introduction to \GPB\ and to this series \citep[Paper I,][]{GPB-I}. In
the second and third papers we report on the structure and its changes
of each of the three extragalactic reference sources, \objectname[]{3C
454.3}, \objectname[]{B2250+194}, and \objectname[]{B2252+172}
\citep[Paper II,][]{GPB-II}, and on the degree of astrometric
stability of the ``core'' of 3C~454.3 in two extragalactic celestial
reference frames \citep[Paper III,][]{GPB-III}.  In the fourth and
fifth papers, we describe our astrometric analysis technique
\citep[Paper IV,][]{GPB-IV} and present our result on the proper
motion, parallax, and orbit of \IM\ \citep[Paper V,][]{GPB-V}. In this
paper (Paper VI), we discuss the locations of the sources of radio
emission in the \IM\ system and give a physical interpretation of the
sources' origins.  In the last paper of our series \citep[Paper
VII,][]{GPB-VII}, we discuss the radio images of \IM\ and include a
movie of this star's changes over the duration of our 8.5~yr observing
program.
 
\IM\ is a close binary with orbital period $\sim$24.65 days and an
essentially circular orbit with an eccentricity of $0.006\pm0.007$
\citep*{BerdyuginaIT1999}.  It is classified by \citet{Hall1976} as an
RS~CVn.  The system is at a distance of $96.4\pm0.7$~pc \citep[Paper
V; see also][]{PerrymanE1997} and has an inclination $65\arcdeg \leq i
\leq 80\arcdeg$ \citep{BerdyuginaIT1999,Lebach+1999}.  The primary is
a K2~III star \citep{BerdyuginaIT1999} which is magnetically active,
showing bright emission features (e.g., \ion{Ca}{2} H and K,
\ion{Mg}{2} H and K, \ion{C}{4}) that are presumably produced by
high-temperature species in its chromosphere and transition region
(\citealt*{HuenemoerderRB1990}; \citealt{Dempsey+1996};
\citealt{Olah+1998}).  In addition, Doppler optical images of the
photosphere of the primary exhibit large, relatively dim ``spot''
regions, covering collectively $>$15\% of the visible stellar surface
\citep{Berdyugina+2000}.  The sun-like secondary is $\sim$60 times
less luminous in the optical than the primary, and has also been
detected spectroscopically \citep{Marsden+2005}.

Radio emission from \IM\ was first detected by
\citet*{SpanglerOH1977}.  Since then, \IM\ has been included in two
radio surveys of RS~CVn systems \citep*{MorrisM1988,DrakeSL1989}, but
few details of its radio properties are published.  Its flux density
at centimeter wavelengths has ranged between $\sim$0.2~mJy and
$\sim$80~mJy \citep[Paper I;][]{Lebach+1999,Boboltz+2003}, and can be
highly variable on sub-hour time scales \citep{Lebach+1999}.  VLBI
observations of \IM\ were made in the early 1990's as part of an
astrometric program to link the Hipparcos optical reference frame to
the extragalactic radio reference frame
\citep{Lestrade+1995,Lestrade+1999}, but no image of \IM\ from this
program was published.

Models for the microwave radio emission of RS~CVn binaries suggest
three possible source regions for the emission within the system: (i)
magnetic-loop structures attached to one stellar component, namely the
active subgiant or giant, in the binary
\citep[e.g.,][]{Mutel+1985,Franciosini+1999}; (ii) a joint
magnetosphere for the two components of the binary
\citep[e.g.,][]{UchidaS1983,Ransom+2002}; and (iii) the region between
the two components \citep{Lestrade1996}.  Multi-epoch astrometric VLBI
observations can potentially distinguish among these scenarios.  In
the related case of the close binary in the Algol system,
\citet{Lestrade+1993} were able, with astrometric VLBI from four
epochs, to identify the cooler K subgiant star, and not its B dwarf
companion or the intermediary region, as the likely source of the
radio emission.  Until now, no such identification has been made for
any RS~CVn system.

A total of 35 additional sessions of astrometric VLBI observations of
\IM\ were conducted between 1997 and 2005 in support of \GPB\@.
Consequently, \IM\ is now more extensively observed at centimeter
wavelengths than any other binary.  In \S\,\ref{known} below, we give
the previously determined orbital parameters of \IM, and set the stage
for a discussion about the source region of the radio emission in the
binary system.  In \S,\ref{reduc}, we give an overview of the VLBI
observations and astrometric analysis procedures.  In
\S\,\ref{results}, we summarize the astrometric results for \IM\
presented in \citetalias{GPB-V}, focusing in particular on the
apparent orbit of the radio emission region and the distribution of
the residuals on the sky.  We discuss our results in \S\,\ref{discuss}
and give our conclusions in \S\,\ref{conclusions}.

\section{Previously Determined Orbital Parameters of IM~Peg \label{known}}

Optical spectroscopic and photometric observations provide accurate
values for most of the basic physical properties and orbital elements
of the \IM\ binary system.  Table~\ref{thestar} summarizes these
results.  The orbits of the primary and secondary stars projected on
the sky are particularly relevant to the problem of determining the
location of the radio emission within the binary system.  If the
emission source is closely tied to either of the two stars, then it
likely travels nearly the same projected orbital path as that star.
However, if the emission source arises primarily in the interbinary
region, the source could remain more nearly stationary near the center
of mass of the binary.

\begin{deluxetable}{l c c@{~~~}c c c}
\tabletypesize{\scriptsize}
\tablecaption{Properties and Previously Determined Orbital Parameters of \IM}
\tablewidth{0pt}
\tablehead{
  \colhead{Parameter} &
  \colhead{} &
  \multicolumn{2}{c}{Value} &
  \colhead{} &
  \colhead{Reference\tablenotemark{a}}
}
\startdata
Trigonometric Parallax (mas)      &  & \multicolumn{2}{c}{$10.33 \pm 0.76$, $10.370 \pm 0.074$} &  & 1,2 \\
Distance (pc)                     &  & \multicolumn{2}{c}{$96.8^{+7.7}_{-6.2}$, $96.4 \pm 0.7$} &  & 1,2 \\
                                      &  &                      &                                   &  & \\
\tableline
\multicolumn{6}{c}{Stellar Properties\tablenotemark{b}} \\
\tableline
                                      &  &                      &                                   &  & \\
Mass ($\Msol$)                        &  & $1.8 \pm 0.2$   & $1.0 \pm 0.1$                          &  & 3,3 \\
Spectral Type                         &  & K2~III          & G~V?\tablenotemark{c}              &  & 4,3 \\
$T_{\rm{eff}}$ (K)                    &  & $4550 \pm 50$   & $5650 \pm 200$\tablenotemark{c}        &  & 4,3 \\
Radius ($\Rsol$)                      &  & $13.3 \pm 0.6$  & $1.00 \pm 0.07$\tablenotemark{c}       &  & 4,3 \\
Radius (mas)\tablenotemark{d}         &  & $0.64 \pm 0.03$ & $0.048 \pm 0.004$\tablenotemark{c}     &  & 4,3 \\
                                      &  &                      &                                   &  & \\
\tableline
\multicolumn{6}{c}{Orbital Elements\tablenotemark{b}} \\
\tableline
                                      &  &                      &                                   &  & \\
$a \sin i$ ($\Rsol$)                  &  & $16.70 \pm 0.02$     & $30.34 \pm 0.03$                  &  & 3,3 \\
$a \sin i$ (mas)\tablenotemark{d}     &  & $0.806$              & $1.464$                           &  & \\
$P$ (days)                            &  & \multicolumn{2}{c}{$24.64877 \pm 0.00003$}               &  & 3 \\
$i$ ($\arcdeg$)                       &  & \multicolumn{2}{c}{$65$...$80$, $>$55}                   &  & 4,5 \\
$e$                                   &  & \multicolumn{2}{c}{$0.0$ (assumed)}                      &  & 4 \\
$T_{\rm conj}$ (HJD)\tablenotemark{e}     &  & \multicolumn{2}{c}{$2450342.905 \pm 0.004$}              &  & 3 \\
\enddata
\tablenotetext{a}{First reference is for the first entry, second
reference, if present, is for the second entry.}
\tablenotetext{b}{Two entries for lines 3--9 correspond to the two
stars of the binary system, with entries for the primary listed
first.}
\tablenotetext{c}{The spectral type, effective temperature, and radius
of the secondary are inferred from the flux ratios (at two
wavelengths) of the two stellar components and the values for the
radius and effective temperature of the primary under the assumption
that the secondary is a main sequence star.}
\tablenotetext{d}{Computed for a system distance of $96.4 \pm 0.7$~pc.
The uncertainty in the $a \sin i$ value in $\Rsol$ units is not
propagated into mas, since the uncertainty in the inclination is the
dominant source of error in any spectroscopic determination of the
semimajor axis.}
\tablenotetext{e}{Heliocentric time of conjunction with the K2~III
primary behind the secondary.}
\tablerefs{
  1. {\em Hipparcos} Catalogue \citep{PerrymanE1997};\phn
  2. VLBI \citepalias{GPB-V};\phn
  3. \citet{Marsden+2005};\phn
  4. \citet{BerdyuginaIT1999} ($e=0.006\pm0.007$);\phn
  5. \citet{Lebach+1999}.
  }
\label{thestar}
\end{deluxetable}

The rotation of at least the K2~III primary is synchronized with the
star's orbit \citep{Olah+1998}, which is expected to be circular based
on tidal theory \citep{Zahn1977}, and indeed determined to be very
nearly circular \citep{Olah+1998,BerdyuginaIT1999,Marsden+2005}.
Projecting a circular orbit with system inclination $i \gtrsim
55\arcdeg$ (see Table~\ref{thestar}) on the sky yields a highly
eccentric, elliptically-shaped orbit.  The semimajor axis lengths of
the projected (elliptical) orbits of the primary and secondary, $a_1$
and $a_2$, respectively, are equal to those of the true (circular)
orbits, which are constrained by spectroscopic data.  Using the values
from such data for $a_1 \sin i$ and $a_2 \sin i$ in
Table~\ref{thestar}, the narrow range of allowed inclination values
$65\arcdeg \leq i \leq 80\arcdeg$ \citep{BerdyuginaIT1999}, and a
system distance of $96.4 \pm 0.7$~pc \citepalias{GPB-V}, we deduce
that the semimajor axes of the orbits for the primary and secondary in
angular units are $a_1 = 0.85 \pm 0.04$~mas and $a_2 = 1.55 \pm
0.06$~mas.  These values imply that the maximum dimension of the
projected orbit of each star is large enough to be detected with
astrometric VLBI and that the projected orbits of the primary and
secondary stars are clearly distinguishable.  If the radio emission
from \IM\ is indeed spatially associated largely or entirely with one
of the two stars of the binary, then a projected orbit derived from
our astrometric observations can distinguish between these two
possibilities.

\section{Overview of Observations and Astrometric Analysis Procedures \label{reduc}}

\subsection{VLBI Observations \label{obs}}

Our 35 sessions of 8.4~GHz ($\lambda = 3.6$~cm) VLBI observations each
used a global VLBI array of between 12 and 16 telescopes.  For each
session, we interleaved observations of \IM\ with either two or three
extragalactic reference sources nearby on the sky, so that we could
employ the phase-referencing technique
\citep[e.g.,][]{Shapiro+1979,Bartel+1986,Lestrade+1990,BeasleyC1995}
and determine an accurate astrometric position for \IM\@.  For a full
description of our array, typical observing schedule, and data
recording parameters, see \citetalias{GPB-II}.  Other aspects of our
observing strategy, e.g., seasonal and orbit-phase distributions of
our observation sessions, are discussed in \citetalias{GPB-V} and
\citetalias{GPB-VII}.  Since the locations of the sources of radio
emission in \IM\ are of special interest in this paper, we emphasize
here that we took care in our scheduling to achieve an approximately
even distribution of orbital phase, without strong correlations
between phase and either year or season.

\subsection{The Astrometric Technique \label{technique}}

The phase-referencing process by which we estimated the position of
\IM\ for each observing session included both phase-delay fitting with
a Kalman-filter estimator and phase-reference mapping (see Paper IV
for details).  This process improves our astrometric accuracy by
allowing us to effectively model the contributions of the troposphere,
ionosphere, and the extended structure of \IM's radio emission regions
(see \S\,\ref{thechoice}), in spite of the often low flux density (as
low as $\sim$0.2~mJy) of the stellar radio emission.  The final image
of \IM\ produced for each session is referenced to the
quasi-stationary component C1, the 8.4 GHz ``core,'' of 3C~454.3.  The
identification and stationarity of this reference point are discussed
in Papers II and III, respectively.

\subsection{Choosing a Position for \IM\ at Each Epoch \label{thechoice}}

As mentioned in \citetalias{GPB-V}, our phase-referenced images reveal
three general categories for the radio source structure of \IM: (1)
single-peaked with peak located near the center of a marginally
extended source; (2) single-peaked with peak located off center of an
elongated source; and (3) double-peaked (or in one case apparently
triple-peaked) with maximum separation between peaks of $\sim$1.5~mas.
An example from each of these categories is given in
\citetalias{GPB-V}.

For the definition of the ``observed'' position for \IM\ we considered
three possible choices: (i) the position of the brightness peak,
interpolated between pixels, of the image for each epoch, (ii) the
central position of the single elliptical Gaussian fit to the image
for each epoch, and (iii) the central position of the single fit
Gaussian for each single-peaked epoch and the position of the
unweighted midpoint between interpolated brightness peaks for each
multiple-peaked epoch.

We then fit the astrometric model described in \S\,\ref{model} below
to each set of positions.  We obtained the best fit for choice (iii):
the chi-square per degree of freedom for the resulting weighted
least-squares fit was 30\% lower than for the worst fit, that for
choice (i).  We therefore adopted set (iii) as our standard set of 35
\IM\ positions for the estimate of position at epoch, proper motion,
parallax, and orbital parameters of \IM. Set (iii) represents in
effect smoothed data, taking into account the extended structure of
the source.  For the study of the locations of the sources of radio
emission in this paper, however, the distribution of the locations of
emission peaks is more important, and therefore set (i) is used for
such an investigation, as described below in \S\,\ref{skymodel}.

\subsection{The Astrometric Model \label{model}}

We fit to the 35 positions of set (iii) a nine-parameter model
\citepalias[see also][]{GPB-V} describing the motion of the \IM\ radio
emission on the sky.  The model parameters are the five scalar
parameters needed to specify \IM's position at a reference epoch, its
proper motion, and its parallax, plus four more to specify the
projection on the sky of its (assumed) zero-eccentricity orbit of
known period.  To maintain the linearity of the model in all unknown
parameters, we let the orbit parameters for each coordinate ($\alpha$
and $\delta$) be the amplitudes of the sine ($A_{s\alpha}$,
$A_{s\delta}$) and cosine ($A_{c\alpha}$, $A_{c\delta}$) terms in
orbital phase.

\section{Results \label{results}}

\subsection{Astrometric Solution \label{solution}}

\begin{figure}
\epsscale{0.75}
\plotone{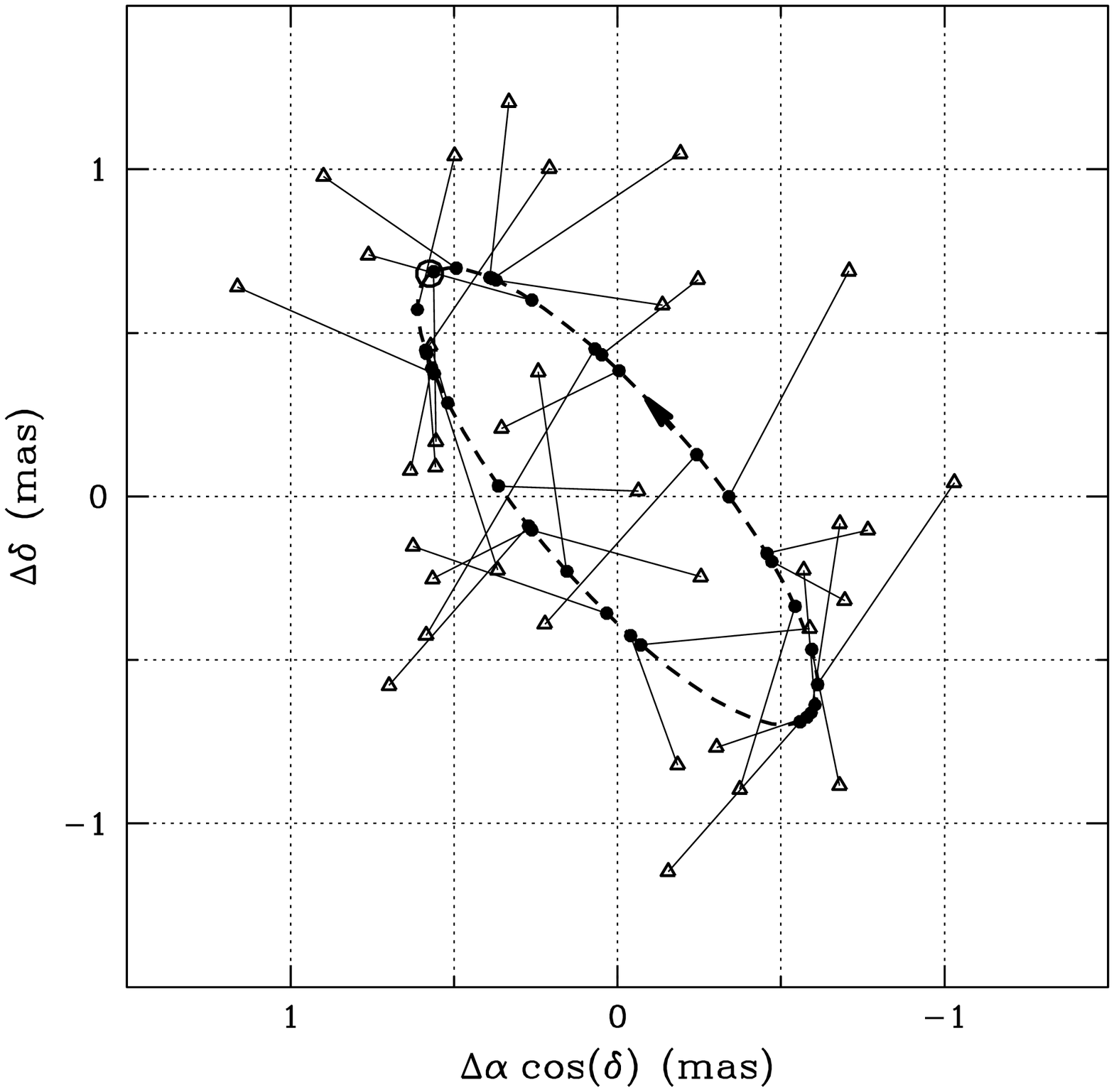}
\caption{The orbit (dashed ellipse) derived
from the nine-parameter fit to the set of 35 positions of the \IM\
radio source.  The inferred direction of motion on the sky is
counter-clockwise, as indicated by the arrow on the ellipse. The
ascending node is given as an open circle on the orbit in the
northeast.  In other words, the part of the ellipse with the arrow on
it is closest to Earth. The observed position for each epoch is
plotted (with an open triangle) after subtraction of the estimated
position at reference epoch, proper motion, and parallax.  This
position corresponds to the peak of a single Gaussian component fit to
the source region or the midpoint between two (or three) local maxima
in the source region.  A solid line connects each observed position
with the corresponding position indicated by a dot on the estimated
orbit (see text).}
\label{IMorbit}
\end{figure}

\begin{figure}
\epsscale{0.75}
\plotone{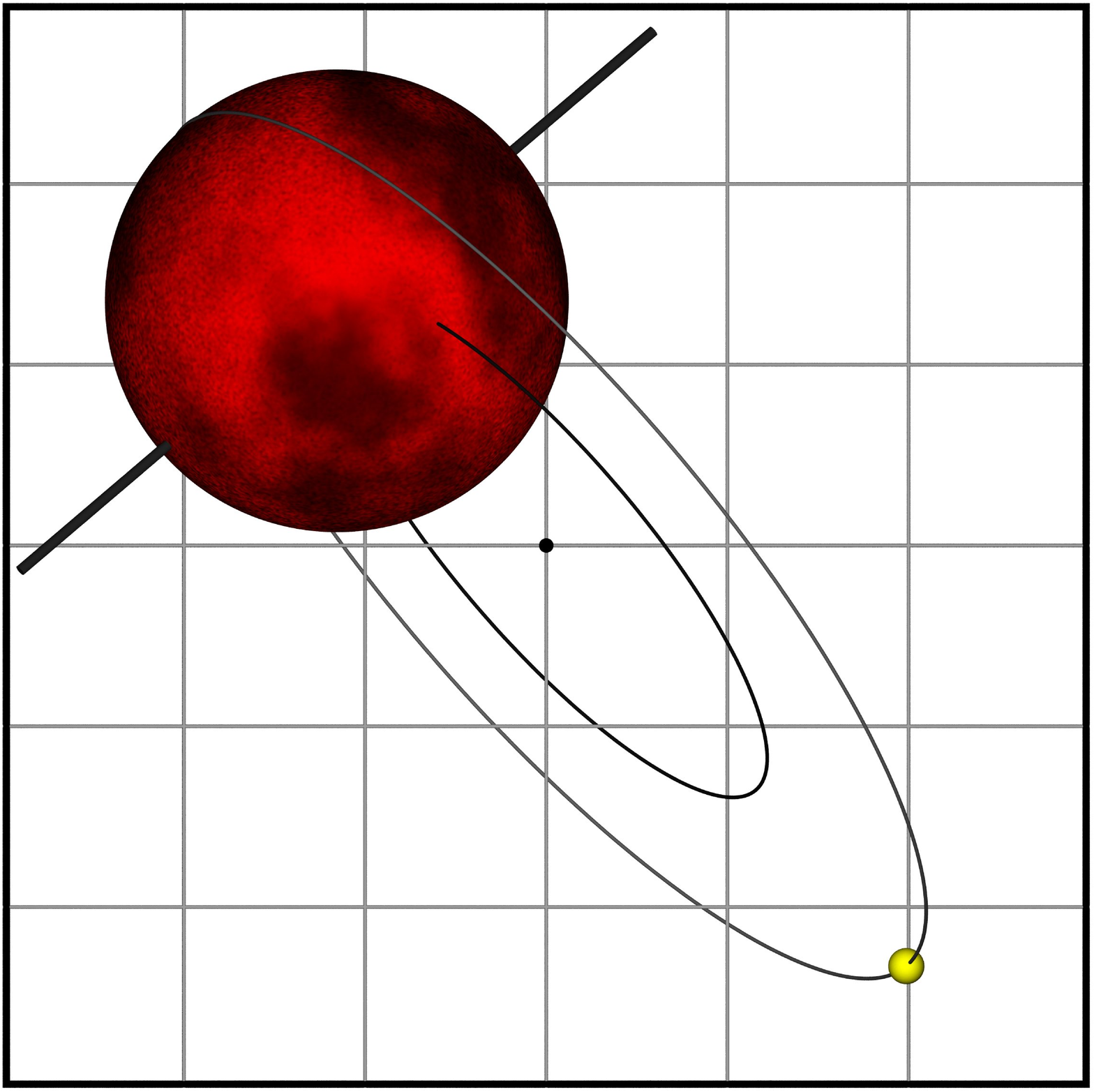}
\caption{Artist's three-dimensional rendering of the \IM\ binary
system as seen from Earth.  Grid lines are drawn every 0.5~mas.  The
primary is the larger red star with dark spots, while the secondary is
the smaller yellow star.  The projected orbit of the primary is the
same as the inferred radio source orbit shown in Figure~\ref{IMorbit}.
The size of the orbit of the secondary is computed from that of the
radio orbit and the ratio of the component masses given in
Table~\ref{thestar}.  The diameters of the primary and secondary stars
correspond to the nominal values given in Table~\ref{thestar}.  The
black dot at the center of the figure represents the center of mass of
the binary system.  The system is shown with the primary at its
ascending node.  The spin axis of the primary is indicated by the rod
passing through the star's center.  It is oriented parallel to the
orbit normal (see \S\,\ref{discuss}).  Note that the visible pole of
the primary is near the south-east side of the stellar disk.}
\label{IMschem}
\end{figure}

The full astrometric solution is presented in Table 3 of
\citetalias{GPB-V}.  In this paper we focus on the orbit.  For the
convenience of the reader and for easier comparison with previously
determined orbital parameters, we reproduce in Table~\ref{fitresults}
the values for the four orbit parameters determined in
\citetalias{GPB-V}.  The estimated orbit and the 35 residual position
determinations with their corresponding orbit-model-predicted
positions are plotted in Figure~\ref{IMorbit}.  The inferred direction
of motion is counter-clockwise as indicated by the arrow.  The size,
shape, orientation, and ``timing'' of the orbit do not change
significantly if we choose instead set (i) or set (ii) of astrometric
positions described in \S\,\ref{thechoice}.  The orbit is also robust
against reasonable changes in either the set of epochs included in the
fit or the addition to the astrometric model of a constant proper
acceleration.  A full discussion of our error analysis is given in
\citetalias{GPB-V}. The orbit of the secondary is obtained from the
ratio of the component masses given in Table~\ref{thestar}. In
Figure~\ref{IMschem} we show an artist's three-dimensional rendition
of \IM\ with the primary and secondary in their estimated orbits as
seen from Earth.

\begin{deluxetable}{l c c}
\tabletypesize{\scriptsize}
\tablecaption{Orbit Parameters of \IM\ Radio Source}
\tablewidth{0pt}
\tablehead{
  \colhead{Parameter} &
  \colhead{Estimate} &
  \colhead{Standard Error\tablenotemark{a}}
}
\startdata
\multicolumn{3}{c}{--- The parameters of the linear orbit model: ---} \\
$A_{s\alpha}$ (mas)                   & $-0.59$         & 0.10 \\
$A_{s\delta}$ (mas)                   & $-0.66$         & 0.11 \\
$A_{c\alpha}$ (mas)                   & \phm{$-$}$0.15$ & 0.09 \\
$A_{c\delta}$ (mas)                   & $-0.23$         & 0.11 \\
\multicolumn{3}{c}{--- The equivalent values of the more commonly used orbit parameters: ---} \\
$a$ (mas)                             & 0.89            & 0.09 \\
Axial ratio\tablenotemark{b}          & 0.30            & 0.13 \\
$\Omega(\arcdeg$)\tablenotemark{c}    & 40.5            & 8.6 \\
$T_{\rm conj}$ (JD)                   & 2450342.56      & 0.44 \\
\enddata
\tablenotetext{a}{See \citetalias{GPB-V} for an explanation of our errors.}
\tablenotetext{b}{The axial ratio, i.e., the ratio of the minor to the
major axis, is equal to the absolute value of the cosine of the
inclination for our zero-eccentricity orbit. The axial ratio therefore
corresponds to an inclination of $73\pm8$\arcdeg.}
\tablenotetext{c}{Position angle (east of north) of the ascending
node.  We follow the convention of identifying the ascending node as
the one at which the source is receding from us as it passes
through the plane of the sky.}
\label{fitresults}
\end{deluxetable}

We emphasize three points from Table~\ref{fitresults}, each of
importance for our study of the location of the sources of radio
emission:

\begin{trivlist}
\item{1.} Our estimate of the length of the semimajor axis of $0.89
\pm 0.09$~mas \citepalias{GPB-V} is both statistically significant and
consistent within $1\sigma$ with that of the semimajor axis inferred
for the \IM\ primary from optical spectroscopy (see \S\,\ref{known},
Table~\ref{thestar}).

\item{2.} The ratio of the length of the minor axis to that of the
major axis of the projected orbit of $0.30\pm0.13$ is relatively small
allowing for a relatively accurate estimate of the position angle
(p.a.) of the ascending node, $\Omega = 40.5\arcdeg \pm 8.6\arcdeg$,
and hence the p.a.\@ of the projected orbit normal, $\Omega -
90\arcdeg = -49.5\arcdeg \pm 8.6\arcdeg$.

\item{3.} The time of conjunction, $T_{\rm conj} = 2450342.56\pm0.44$
JD, implied by our orbit is also consistent within $1\sigma$ with that
derived for the primary from optical spectroscopy (see
Table~\ref{thestar}).

\end{trivlist}

\subsection{The Mean Location of the Source of the Radio Emission \label{meanlocation}}

In \citetalias{GPB-V} we show that the residuals of our nine-parameter
astrometric fit to set (iii) of the 35 positions of \IM\ scatter about
a well determined orbit on the sky.  How does this fit determine the
mean location of the radio emission source with respect to the optical
primary and secondary of the binary system?  With a near
zero-eccentricity orbit, the axial ratio of the projected radio orbit
corresponds to an inclination of $73\arcdeg \pm 8\arcdeg$.  Combining
our value for the inclination with the $a \sin i$ estimate of
\citet{Marsden+2005} for each stellar component (see
Table~\ref{thestar}) leads to semimajor axes of $0.84 \pm 0.03$~mas
for the primary and $1.53 \pm 0.06$~mas for the secondary.  Thus our
estimated semimajor axis of the radio orbit of $0.89\pm0.09$~mas
agrees well with that of the spectroscopic orbit of the primary, but
differs significantly from that of the secondary.  Furthermore, the
time of conjunction, $T_{\rm conj}$, obtained for the radio orbit is
only $0.3 \pm 0.4$~d earlier than that found by \citet{Marsden+2005}
(see Table~\ref{thestar}).  The estimated radio orbit is thus in the
same phase within the error as the spectroscopic orbit.  The above
difference in $T_{\rm conj}$ corresponds to a physical offset between
the center of the primary and the mean position of the radio emission
from our model orbit that is only $0.12 \pm 0.15$ times the radius of
the primary and is not significantly different from
zero.\footnote{This result depends on the assumption that there is no
offset between the center of the primary and the mean position of the
radio emission that is constant or steadily increasing/decreasing over
the 8.5 years of VLBI observations. Such an offset could of course not
be determined in our fit since it would be absorbed in our position
and proper-motion estimates.  However, as we demonstrate in
\S\,\ref{skymodel}, the distribution of our position solutions well
covers the disk of the primary.  Therefore it appears that the
emission locations are very closely linked to the primary and any
constant or linearly changing offset of the center of this
distribution from the center of the primary is likely smaller than the
radius of the primary \citepalias[see also][]{GPB-V}.}  That is, the
offset is with $1\sigma$ accuracy likely to be distant from the center
by less than $27$\% of the radius of the primary. We therefore have
strong observational evidence to conclude that the active primary is
the source of the radio emission in \IM.  This result makes \IM\ only
the second close binary system for which such an identification has
unambiguously been made, the other being the close binary in the Algol
system \citep{Lestrade+1993}.

\subsection{Distribution of Position Residuals on the Sky \label{skymodel}}

\begin{figure}
\epsscale{1.0}
\plotone{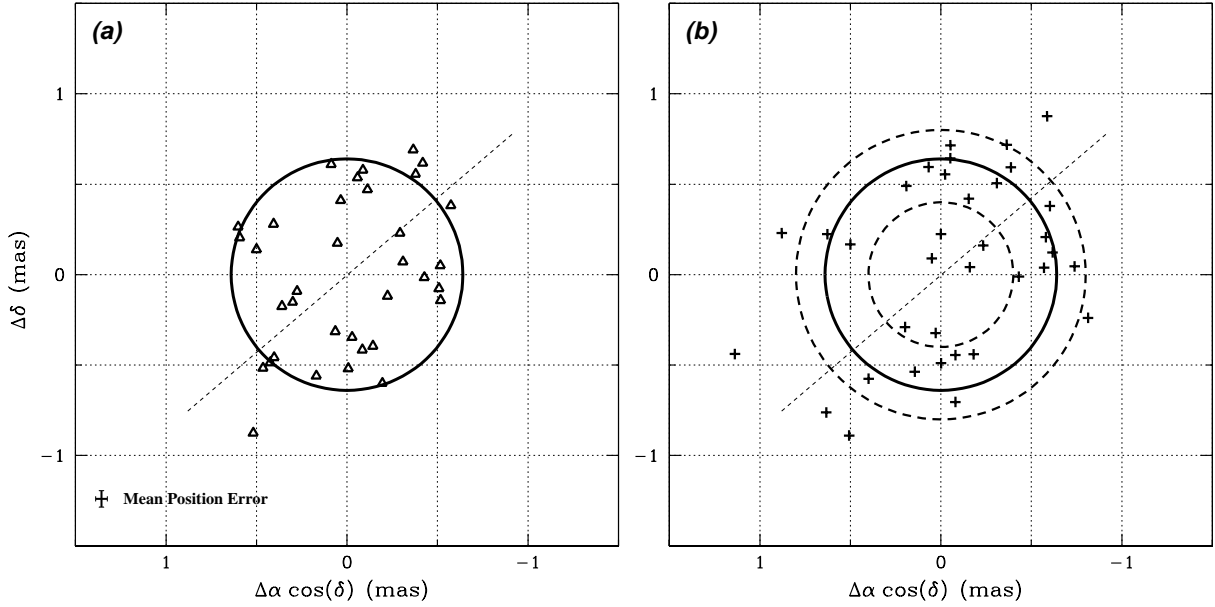}
\caption{$(a)$ Sky distribution of the observed positions plotted in
Figure~\ref{IMorbit} (open triangles) after removal of the estimated
orbit. The errors bars shown in the bottom-left-hand corner represent
the mean position error for the 35 epochs (see text).  $(b)$ As in
$(a)$ but now the sky distribution of the brightness-peak positions
(plusses). To plot these positions, we added to the position in $(a)$
for each epoch the difference between the position of the brightness
peak (set i) and the position of the Gaussian or midpoint (set iii --
see text).  The mean position error in $(b)$ is virtually identical to
that shown in $(a)$.  The solid circle in each panel indicates the
disk of the primary with a radius of 0.64 mas and centered at the
origin by assuming that the center of the primary follows the
estimated orbit.  The dashed line passing through the center of each
panel indicates the projected spin axis of the primary, assuming it is
parallel to the orbit normal (see \S\,\ref{discuss}).  The dashed
circles in $(b)$ separate the three regions used for the counting
analysis described in \S\,\ref{reducspill}.}
\label{residonsky}
\end{figure}

We show in Figure~\ref{residonsky}$a$ the residuals of our set (iii)
position solutions to our nine-parameter weighted least-squares
astrometric fit.  The residuals correspond to the positions plotted in
Figure~\ref{IMorbit} after removal of the model orbit. In addition we
plot the disk of the primary, placing its center at the origin of the
diagram.  There are two important features in the sky-distribution of
the residuals: First, almost all of the residuals lie inside of the
disk of the primary, with some residuals going only slightly beyond
it.  In fact, the 0.55~mas root-mean-square (rms) scatter (0.35~mas in
$\alpha$ and 0.46~mas in $\delta$) is almost equal to the angular
radius of the primary of $0.64\pm0.03$~mas (Table~\ref{thestar}) but
smaller than the semimajor axis of the orbit of the primary of
$0.89\pm0.09$~mas (Table~\ref{fitresults}) .  With a relatively small
mean standard error in the position determination of the stellar radio
source of 0.07~mas in $\alpha$ and 0.09~mas in $\delta$, the scatter
is largely intrinsic to the emission source.\footnote{The mean
standard error in the position determinations is the root-sum-square
of the mean statistical standard error associated with determining the
position of the reference point in the image of \IM\ at each epoch and
an estimated $\sim$0.06~mas astrometric error in each \IM\ coordinate
\citepalias{GPB-V}.  The astrometric error includes the rms 'jitter'
of 3C~454.3 component C1 relative to our extragalactic reference frame
\citepalias[see][]{GPB-III}.}  Second, the residuals are scattered
preferentially along a northwest-southeast oriented axis,
approximately along the projected normal of the orbit. It is likely
that this preferential elongation of the distribution of the residuals
is also intrinsic to the emission source.  The synthesized
interferometer beam is by comparison more nearly oriented north-south
(mean $\rm{p.a.} \sim -7\arcdeg$).  Moreover, when we fit for the
purposes of error analysis the same nine-parameter model to the
positions of the secondary reference source B2250+194 (which is
$\sim$5 times farther away from 3C~454.3 than \IM), we found an
approximately threefold smaller scatter in the rms residuals and a
smaller correlation between coordinates.

For the physical interpretation of the location of the source of radio
emission it is more meaningful to study the distribution of the
positions of the brightness peaks (set i) which indicate where the
dominant part of the emission originates.  We therefore added to the
residual for each epoch the difference between the position of the
interpolated brightness peak (from set i) and the position of the
Gaussian or midpoint (from set iii) used in our astrometric fit.  Use
of these ``modified'' residuals (hereafter referred to simply as
residuals) ensures that we are plotting the offsets of the
interpolated brightness peak at each epoch from our estimated orbit
(see Figure~\ref{residonsky}$b$). The rms scatter about the mean is
$0.62\pm0.03$~mas, and, as expected, somewhat larger than the rms
scatter of 0.55 mas in Figure~\ref{residonsky}$a$.

We determined the 0.03~mas standard error for the above sample
estimate of rms scatter using a bootstrap method
\citep[see][]{EfronT1993}.  More specifically, we regarded the 35
residuals as the parent distribution and chose from this distribution
a new set of 35 residuals, with each one being randomly selected from
the parent distribution (with replacement), until we obtained 500
different sets of 35 residuals.  For our estimate of the standard
error of the rms scatter of the true distribution, i.e., before
sampling by our observations, we take the standard deviation of the
500 values of rms scatter that we computed from the 500 sets.  We used
this bootstrap method to derive standard errors for all parameters
estimated from, or compared to, the 35 residuals, both here and, where
relevant, hereafter.

\subsection{Simulation of the Distribution of Position Residuals on the Sky and Comparison with Observations \label{simulations}}

How can the scatter of the locations of brightness peaks be
interpreted in terms of the geometry of the orbit and the relation to
the primary?  In other words, how closely is the seemingly preferred
direction of the scatter of these locations related to the normal of
the orbit and the spin axis of the primary, and how far from the
surface of the primary do the brightness peaks originate?  Motivated
by our astrometric solution and the distribution of the brightness
peak locations in Figure~\ref{residonsky}$b$, we constructed a
three-dimensional model to simulate emission regions in the corona of
the primary.  We call this model the coronal emission model (CEM). We
then considered different distributions of locations of emission
regions to find the best match to our observations and thereby obtain
a reasonable model for the location of the source of radio emission in
\IM.
 
\subsubsection{Latitude-independent Coronal Emission Model with Spillover Emission \label{basic}}

We start with the latitude independent version of the CEM. In this
model we assume that on average the radio emission is centered, in
projection on the sky, on the center of the stellar disk, and allow
the location of peak brightness to fall with equal probability above
any point on the stellar surface.  We allow emission to occur at any
height above the photosphere of the star, but assume that the
probability that the emission peaks at a given height decreases
exponentially with scale height, $H$.\footnote{We are not asserting
here that an exponentially decreasing emission probability is
physically realistic.  An exponential function provides via a single
parameter an analytical means of estimating the statistical
distribution of emission heights.  Other functions (e.g., uniform
emission probability) employ a sharp cut-off at an arbitrary height
and, more importantly, were not able to reproduce in projection the
full extent and distribution (in the three regions described in
\S\,\ref{reducspill}) of the observed residuals.}  We then project the
CEM onto the sky to compare our model distribution to the distribution
of the locations of brightness peaks as plotted in
Figure~\ref{residonsky}$b$. For emission from locations not occulted
by the disk of the star, we simply project that location onto the sky.
For emission locations occulted by the disk of the star, we allow
``spillover;'' i.e., we move the predicted location of observed peak
emission outward along the radial line on the sky to a location we can
see slightly beyond the edge of the stellar disk (see below).
Allowing for spillover is perhaps reasonable, since we had a 100\%
detection rate and observed extended radio structure at every epoch;
however, as we shall later show, nature does not seem to have followed
this path and there are other ways to understand the 100\% detection
rate.  We do not account in any sophisticated way for the transmission
of spillover emission through the corona, e.g., by considering
scattering, absorption, and/or refraction along the line of
sight. Instead, we arbitrarily allow a small radial distribution of
this emission on the sky spanning a projected radius of 1.0--1.1 times
the stellar radius.  The adjustable parameter in the
latitude-independent CEM is $H$.

\begin{figure}
\epsscale{0.80}
\plotone{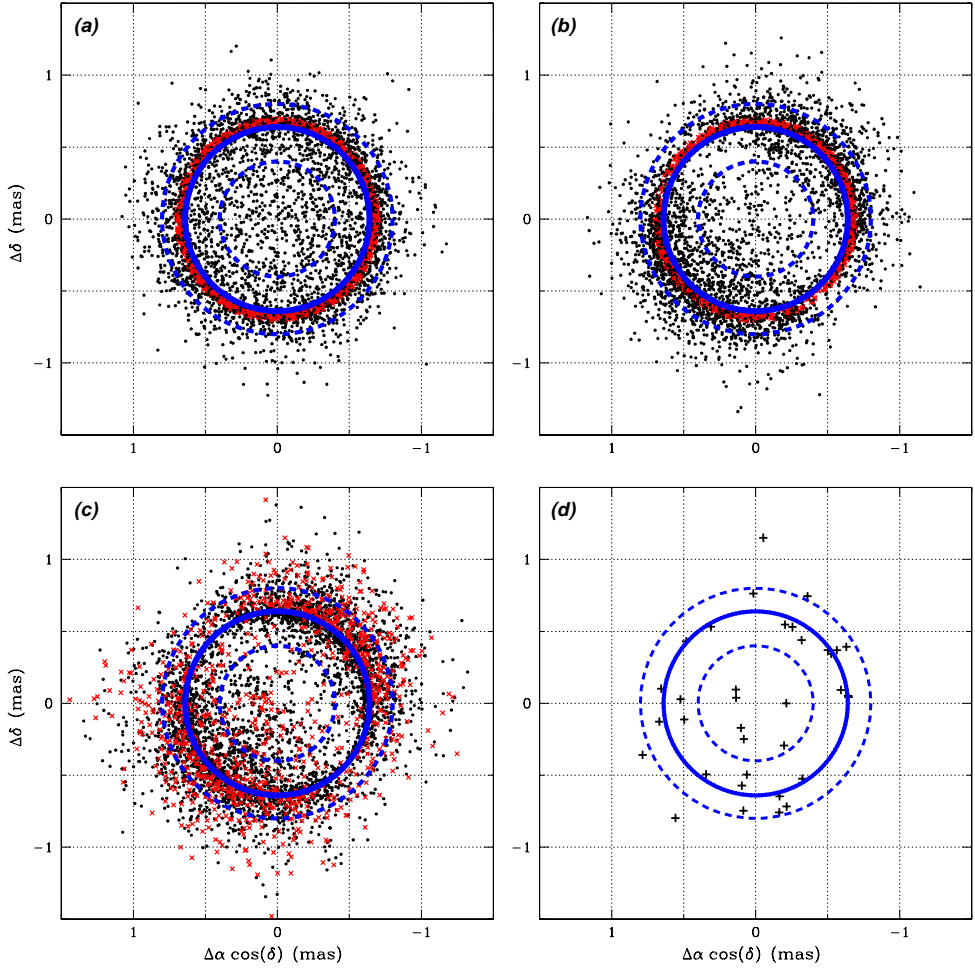}
\caption{\small $(a)$ Sky projection of 5000 random emission points
from the latitude-independent CEM with spillover emission described in
\S\,\ref{basic}.  The scale height in the model is $0.17$ times the
0.64~mas stellar radius, a value which produced the best match to the
observed distribution.  The dots represent emission points that are
not occulted by the disk of the star.  The ``x's'' (red in the colored
version) just above the stellar disk represent spillover emission from
points occulted by the disk of the star (see text). $(b)$ As in $(a)$
but now the points from the latitude-dependent CEM with spillover
emission described in \S\,\ref{preflat}.  The scale height is $0.15$
times the 0.64~mas stellar radius, again chosen so as to produce the
best match to the observations. $(c)$ As in $(a)$ but now the points
from the latitude-dependent CEM with reduced spillover emission
described in \S\,\ref{reducspill}.  The scale height was 0.20 times
the 0.64~mas stellar radius (again chosen for the best match).  $(d)$
Sky projection of 35 randomly selected emission points (plusses) from
the latitude-dependent CEM with reduced spillover emission.  For all
panels, the solid circle gives the disk of the primary and the dashed
circles separate the three regions used for the counting analysis
described in \S\,\ref{reducspill}.}
\label{CEMplot}
\end{figure}

To execute the comparison, we computed several sets of 5000 locations
each for the CEM, each set with the nominal value for the stellar
radius but a different $H$.  For each set we compared the rms of the
distribution of the radio-emission locations to the $0.62 \pm
0.03$~mas rms scatter of the observed residuals.  We found agreement
between the two rms values for $H = 0.11 \pm 0.03$~mas; i.e., a scale
height only $0.17 \pm 0.05$ times the 0.64~mas stellar radius. The
standard error in $H$ is the variation in that parameter needed to
produce an increase or decrease in rms of the distribution of the
radio-emission locations equal to the standard error ($\pm 0.03$~mas)
of the rms of the scatter in the residuals.  We provide an
illustration of the latitude-independent CEM in
Figure~\ref{CEMplot}$a$.

The figure illustrates that the latitude-independent CEM produces a
circularly symmetric distribution of emission locations on the
sky. Such a distribution fails qualitatively to describe the
elongation in the scatter of the residuals.  For a more direct
comparison with the scatter of the 35 residuals, we produced several
sets of 35 (random) emission locations from the latitude-independent
CEM, and for none of them was there more than a 16\% difference
between the rms along the line at which it was greatest from that
along the line for which it was least.  In contrast, the residuals
show a $52\% \pm 22\%$ difference between the rms scatter along the
line for which this rms scatter is maximum ($\rm{p.a.} = -38\arcdeg
\pm 8\arcdeg$) and the line for which this scatter is minimum.

\subsubsection{Latitude-dependent Coronal Emission Model with Spillover Emission \label{preflat}}

To improve upon our CEM, we maintained our assumptions concerning
spillover, but dropped our assumption of equal probability density per
unit surface area.  Instead, we allowed this probability density to
vary with stellar latitude, $\lambda$.  This enhancement of our model
was motivated by Doppler optical surface images of the \IM\ primary
which show persistent ($\sim$1--3 years and possibly longer),
high-intensity spot features in both the mid-latitude range and
directly over the visible pole
\citep{Berdyugina+2000,BerdyuginaM2006,Marsden+2007}.  Given the close
alignment expected for the spin axis and the orbit normal (see
\S\,\ref{discuss}), we assumed that the spin axis of the primary is
inclined $73\arcdeg$ to our line of sight, an angle equal to our
VLBI-derived estimate of the inclination of the orbit (see
Table~\ref{fitresults}).  We tried a number of different functional
forms for the latitude dependence of the probability density
distribution.  We found that a distribution proportional to $k +
\sin|\lambda|$, with $k = 0.14^{+0.11}_{-0.04}$, yielded a good fit to
the sky distribution of the residuals.  For this $k$ value, the mean
probability density per unit surface area for emission near each pole
($|\lambda| \geq 70\arcdeg$) is $3.6^{+0.4}_{-0.7}$ higher than that
for emission near the equator ($|\lambda| \leq 20\arcdeg$).  The
indicated $\sim$70\% confidence limits reflect the uncertainty of the
elongation of the scatter of the residuals, as estimated with our
bootstrap approach.

We also varied the scale height, and found best agreement between the
distribution of the CEM and the residuals for a value $0.15 \pm 0.05$
times the stellar radius.  In Figure~\ref{CEMplot}$b$ we provide an
illustration of the latitude-dependent CEM oriented on the sky with
the spin axis lying along $\rm{p.a.} = -38\arcdeg$, to align with the
axis of the elongation of the distribution of our 35 residuals.

\subsubsection{Latitude-dependent Coronal Emission Model with Reduced Spillover Emission \label{reducspill}}

The latitude-dependent CEM with spillover emission gives us a
distribution of projected emission points which is consistent with the
distribution of residuals both in overall extent and degree of
elongation.  To compare the distributions in more detail, we looked at
the number of emission points and residuals in each of three regions
at increasing radial distances from the center of the
distribution. Based on the extent of the distribution, we chose the
following regions: (1) $<$0.40 mas, (2) 0.40--0.80 mas, and (3)
$>$0.80 mas. The regions are indicated by dashed circles in
Figure~\ref{residonsky}$b$ and in each of the panels in
Figure~\ref{CEMplot}.  The results, given in Table~\ref{CEMcompare},
show that the number of emission locations in region 2 is
significantly higher for the latitude-dependent CEM with spillover
emission than for the observed residuals.  Figure~\ref{CEMplot}$b$
shows clearly that, for the latitude-dependent CEM with spillover
emission, many of the emission locations in region 2 are from
spillover.  Indeed, 35\% of emission locations in this region, and
28\% of all points in this model are due to spillover.  The percentage
of instances for which the radio brightness peak arises from spillover
emission may, however, be much lower.  Here is why. In nine of 35
epochs, our images of \IM\ show two (or three) local brightness
maxima, with the maxima separated in four of those epochs by $\geq$0.5
mas \citepalias[see][]{GPB-VII}.  Since simultaneous emission from
multiple regions appears common for \IM, it is reasonable to expect
that spillover emission be dominated in many instances by emission
from the Earth-facing side of the primary.  To investigate this
possibility, we modified the selection of projected emission locations
in the latitude dependent CEM with spillover emission.  Instead of
allowing each occulted location to become spillover emission, we
allowed, first, only 50\% of occulted locations to become spillover
emission, replacing the other 50\% with (randomly and independently
drawn) non-occulted locations.  We then decreased the fraction of
allowed occulted locations to zero in intervals of 5\% (i.e., 45\%,
40\%, \dots), adding the deleted ones at non-occulted locations.  For
each such fraction, we adjusted the scale height, $H$, so as to
maximize the agreement between the distribution of the CEM emission
locations and that of the residuals.  We found the best agreement
between the distribution of emission locations in the CEM and that of
the residuals, considering the number counts in the three regions
defined above, for a reduced frequency of spillover emission with a
fraction of $10\% \pm 10\%$ of allowed locations (see
Table~\ref{CEMcompare}), and a scale height $0.20 \pm 0.05$ times the
stellar radius.  We provide in Figure~\ref{CEMplot}$c$ an illustration
of this latitude-dependent CEM, with reduced spillover emission,
oriented on the sky with the spin axis lying along $\rm{p.a.} =
-38\arcdeg$.  Note that for our best fit model, in which the spillover
from $10\% \pm 10\%$ of emission peaks in the occulted region yields
an observed brightness peak, the percentage of all emission peaks
which occur due to spillover is just $3\% \pm 3\%$.  Thus, for a
random sample of 35 emission locations, this value suggests that only
$\leq$2 of the locations arise from spillover emission.  We show one
realization of 35 randomly chosen emission points from the
latitude-dependent CEM with reduced spillover emission in
Figure~\ref{CEMplot}$d$.

\begin{deluxetable}{c r r r}
\tabletypesize{\scriptsize}
\tablecaption{Comparison of the Distribution of Residuals to the Distribution of CEM Emission Points}
\tablewidth{0pt}
\tablehead{
   &
  \colhead{Residuals} &
  \colhead{lat.-dep. CEM with spillover\tablenotemark{a}} &
  \colhead{lat.-dep. CEM with reduced spillover\tablenotemark{b}} \\
   &
  \colhead{35 epochs} &
  \colhead{35 points\tablenotemark{c}} &
  \colhead{35 points\tablenotemark{c}}
}
\startdata
Region 1 &  $6 \pm 2$ ($17\% \pm 6\%$) &  $2 \pm 1$  \phn($6\% \pm 3\%$) &  $5 \pm 2$ ($14\% \pm 6\%$) \\
Region 2 & $22 \pm 3$ ($63\% \pm 9\%$) & $29 \pm 2$     ($83\% \pm 6\%$) & $24 \pm 3$ ($69\% \pm 9\%$) \\
Region 3 &  $7 \pm 3$ ($20\% \pm 9\%$) &  $4 \pm 2$     ($11\% \pm 6\%$) &  $7 \pm 2$ ($20\% \pm 6\%$) 
\enddata
\tablenotetext{a}{{}Latitude-dependent CEM with spillover emission as
discussed in \S\, \ref{preflat}.}
\tablenotetext{b}{{}Latitude-dependent CEM with reduced spillover
emission as discussed in \S\, \ref{reducspill}.  The tabulated results
correspond to a 10\% allowed fraction of spillover points.}
\tablenotetext{c}{The value and standard error given for each region
represent the mean number and standard deviation for 500 realizations
of 35 randomly chosen CEM emission points (see text).}
\label{CEMcompare}
\end{deluxetable}

\section{Discussion \label{discuss}}

Our astrometric result shows that the sources of radio emission are
consistent with their being centered on average on the primary, and
also consistent with theoretical models which propose that the radio
emission in active close binaries is powered and confined by the
magnetic field of the active star \citep[see,
e.g.,][]{Lestrade+1988,Mutel+1998,Franciosini+1999}.  Any constant or
linearly changing offset of the center of the distribution of emission
locations from the center of the disk of the primary over the 8.5
years of our VLBI observations is not necessarily expected to be
zero. First, any latitude-dependent emission model when combined with
shadowing of one of the pole regions due to the inclination of the
spin axis would predict a constant offset.  Second, a systematic
time-dependence of the latitude distribution of spot centers could
arise given the apparent multi-year stellar-activity cycle of the
\IM\@ primary \citep[e.g.,][]{Berdyugina+2000,Zellem+2010}.  With the
distribution of the radio emission locations at least partly linked
(see below) to such spot features, an offset with a nonzero trend from
1997--2005 could plausibly contribute error to our proper-motion
estimate.  However, each of these possible causes would not result in
the offset being larger than the radius of the primary
\citepalias[see][]{GPB-V}. This estimate, together with the
distribution's matching the disk of the primary, makes us believe that
any offset at any time during our observations is indeed smaller than
the radius of the primary.

Our position residuals are scattered preferentially along an axis with
p.a.\ of $-38\arcdeg\pm8\arcdeg$, which, to within the combined
uncertainties, is equal to the p.a.\ of the sky-projected orbit normal
of $-49.5\arcdeg\pm8.6\arcdeg$. Since the orbit normal is expected to
be closely aligned with the spin axis of the primary as for all
synchronous RS~CVn systems \citep[see][]{Stawikowski1994}, the sources
of radio emission appear to be linked to the spin axis of the
primary. Indeed, a comparison of the scatter of the position residuals
with our simulations shows that the probability density per unit
surface area for radio brightness peaks is $3.6^{+0.4}_{-0.7}$ higher
near the poles ($\lambda \geq 70\arcdeg$) than near the equator
($\lambda \leq 20\arcdeg$).  Since Doppler images of \IM\ show the
presence of persistent, high-intensity spot features at the pole
region of the primary \citep{BerdyuginaM2006,Marsden+2007}, our result
provides statistical evidence that the radio emission regions occur at
the same stellar latitudes as do active surface regions.

Our simulations place restrictions on the altitude of the locations of
radio-emitting structures in the corona of the \IM\ primary.  In our
best-fit model, $\sim$2/3 of the brightness peaks arise within an
altitude of just 0.25 times the stellar radius.  Since the stellar
magnetic field is presumably strongest near the surface, this result
may not be surprising.  However, it is different from the pictures
presented for the close binary systems Algol
\citep[e.g.,][]{Lestrade+1988} and UX~Arietis
\citep[e.g.,][]{Franciosini+1999}, which show emission from
magnetic-loop structures with heights greater than one stellar radius.
If emission does occur high on magnetic loops for \IM, then most of
these loops are small in height compared to the radius of the primary.

Our simulations further allowed us to set at 6\% the ($1\sigma$) limit
on the likelihood (at any given epoch) that the brightness peak arises
from spillover.  Although this limit is model-dependent, we think that
the more general conclusion, which is that the observed brightness
peaks arise mostly from emission regions connected to the Earth-facing
side of the primary, is robust.  Spillover could, however, be
responsible for lower-surface-brightness features of the radio
structure.  In many of our epochs, the \IM\@ radio structure is
multi-peaked or at least significantly elongated.  While the highest
peak in these instances is likely to be associated with emission from
the Earth-facing side, spillover could very well contribute to the
overall shape.  In \citetalias{GPB-VII}, we present the full set of
\IM\ images with the outlines of the disk and orbit of the primary
superimposed, and discuss in detail the size and shape of the radio
emission regions.

\section{Conclusions \label{conclusions}}

\noindent Here we summarize our results and give our conclusions:

\begin{trivlist}

\item{1.} The sources of radio emission are on average located near
the center of the disk of the primary, in particular being within
$12 \pm 15$\% of its radius of $0.64\pm0.03$ mas, provided any 8.5-yr
constant or linearly changing offset is (nearly) zero.  There are
theoretical arguments as well as observational evidence that any such
offset is indeed smaller than the radius of the primary.  Thus \IM\@
is the second close binary with such (nearly) unambiguous
identification.

\item{2.} The positions of the sources of observed radio emission are
scattered over an area on the sky slightly larger than the disk of the
primary and preferentially along an axis with p.a.\ $=
-38\arcdeg\pm8\arcdeg$.  This axis is closely aligned with the
sky-projected orbit normal (p.a.\ $= -49.5\arcdeg\pm8.6\arcdeg$) and
expected spin axis of the primary.

\item{3.} Comparison of our observed positions with simulations
suggests that the radio brightness peaks are more likely to occur at
higher stellar latitudes than near the stellar equator, with the
probability density per unit surface area being $3.6^{+0.4}_{-0.7}$
times higher near the poles ($\lambda \geq 70\arcdeg$) than near the
equator ($\lambda \leq 20\arcdeg$).  The radio emission regions
therefore show a dependence on stellar latitude similar to that
exhibited by active regions on the primary's surface seen as dark
spots with optical Doppler imaging.

\item{4.} This comparison also suggests that these brightness peaks
preferentially arise close to the surface of the primary, with
$\sim$2/3 of them located no more than 0.25 stellar radii above the
surface.

\item{5.} This comparison further suggests that the brightness peaks
are mostly associated with emission regions on the Earth-facing side
of the primary, with peaks caused by spillover from emission regions
on the opposite side of the star occurring rarely, if ever.

\end{trivlist}

\acknowledgements

ACKNOWLEDGMENTS.  We thank the anonymous referee for a constructive
review of the paper and for comments helpful in the preparation of the
final manuscript.  This research was primarily supported by NASA,
through a contract with Stanford University to SAO, and a subcontract
from SAO to York University.  The National Radio Astronomy Observatory
(NRAO) is a facility of the National Science Foundation operated under
cooperative agreement by Associated Universities, Inc.  The DSN is
operated by JPL/Caltech, under contract with NASA.  We have made use
of NASA's Astrophysics Data System Abstract Service, developed and
maintained at SAO.  Jeff Cadieux and Julie Tom{\'e} assisted with the
reduction of the VLBI data during their tenures as students at York
University.

\end{document}